\documentclass[%
 reprint,
showkeys,
 amsmath,amssymb,
aip,pop,
]{revtex4-1}

\usepackage{graphicx}
\usepackage{dcolumn}
\usepackage{bm}
\usepackage{sidecap}
\usepackage{enumerate}

\def\apj{ApJ.}
\def\solphys{Sol. Phys.}
\def\aap{A\&A}

\def\jgr{J. Geophys. Res.}

\def\apjl{Astrophys. J. Lett.}
\def\araa{Annual Rev. of Astron. \& Astrophys.}
\def\aapr{The Astron. \& Astrophys. Rev.}
\def\ssr{Space Science Rev.}
\def\mnras{Monthly Notices of the Roy. Astronom. Soc.}

\def\pre{Phys. Rev. E}

\begin{document}

\preprint{APS/123-QED}

\title{Non-linear Tearing of 3D Null Point Current Sheets} 

\author{P. F. Wyper}
 \email{peterw@maths.dundee.ac.uk}
\author{D. I. Pontin}%
 \email{dpontin@maths.dundee.ac.uk}
\affiliation{%
 Division of Mathematics, University of Dundee,
    Dundee, UK} 


\date{\today}

\begin{abstract}
The manner in which the rate of magnetic reconnection scales with the Lundquist number in realistic three-dimensional (3D) geometries is still an unsolved problem. It has been demonstrated that in 2D rapid non-linear tearing allows the reconnection rate to become almost independent of the Lundquist number (the `plasmoid instability'). Here we present the first study of an analogous instability in a fully 3D geometry, defined by a magnetic null point. 
The 3D null current layer is found to be susceptible to an analogous instability, but is marginally more stable than an equivalent 2D Sweet-Parker-like layer. 
Tearing of the sheet creates a thin boundary layer around the separatrix surface, contained within a flux envelope with a hyperbolic structure that mimics a spine-fan topology. Efficient mixing of flux between the two topological domains occurs as the flux rope structures created during the tearing process evolve within this envelope. This leads to a substantial increase in the rate of reconnection between the two domains.
\end{abstract}

\maketitle


\section{Introduction}

Magnetic reconnection is a process in an almost ideal plasma that permits a stressed magnetic field to become restructured, releasing its free energy. Examples of reconnection-related phenomena include solar flares, geomagnetic storms in the Earth's magnetosphere and saw-tooth crashes in tokomaks, \cite[][and references therein]{Zweibel2009}.

The problem of fast reconnection has been under discussion since the inception of the classical Sweet-Parker (SP) \cite{Parker1957,Sweet1958} reconnection model. In the SP model, the current sheet has a length of the order of the system size $L$, and a width $\delta = L/S^{1/2}$, where $S=v_{a}L/\eta$ is the magnetic Lundquist number based on this length scale and $v_{a}$ is the Alfv\'{e}n speed in the inflow region. Developed to explain energy release in solar flares, the reconnection rate in the SP model ($\sim S^{-1/2}$) is orders of magnitude too slow to explain observations, since in the solar corona $S$ can take values as high as {$S \approx 10^{14}$}. Therefore, one requires a mechanism that scales more favourably with $S$. 

Recently, attention has returned to whether the tearing mode \cite{Furth1963} -- initially disregarded as being too slow -- could enhance reconnection in such large systems \cite[e.g.][]{Huang2013,Biskamp1986,Lee1986}. This followed from the realisation that when the classical tearing analysis---which leads to a  weak growth rate in a fixed neutral sheet---is applied to the SP sheet {with length $L$ and width $\delta = L/S^{1/2}$} the instability grows explosively at high Lundquist numbers following scalings given by \cite{Loureiro2007,Bhattacharjee2009,Loureiro2013}
\begin{equation}
k_{max}L \sim S^{3/8}, \quad \gamma_{max}(v_{a}/L) \sim S^{1/4}, \quad \delta_{inner}/\delta \sim S^{-1/8},
\label{eq:scale}
\end{equation}
where $k_{max}$ is the wavenumber of the fastest growing mode, $\gamma_{max}$ is the growth rate of this mode and $\delta_{inner}$ is the width of the resistive inner layer within which the instability grows in the linear phase. Although tearing of SP sheets had been known about for some time \citep[e.g.][]{Biskamp1986,Lee1986}, it is only much more recently that the scaling relationships of the linear phase were properly developed. This tearing occurs above a critical Lundquist number ($S_{c}$) and aspect ratio ($\mathcal{A}=\delta/L$), typically around {$\approx 10^{4}$ and $\approx $ 50--100}, respectively \cite{Loureiro2013}. In the context of {SP-like} sheets, the tearing mode is often referred to as the `plasmoid instability'.

2D MHD simulation studies have now confirmed that if an SP current sheet fulfills the above criteria, the evolution goes through three phases: (i) non-linear quasi-steady reconnection at the slow SP rate; (ii) tearing and inter-island current sheet thinning, rapidly speeding up the overall reconnection rate; (iii) bursty reconnection mediated by a non-linear hierarchy of current sheets. In this final phase the dynamics is governed by magnetic island formation, coalescence and ejection, and when averaged over time the reconnection rate becomes only weakly dependent upon $S$. Beyond MHD, the inter-island current sheet thinning associated with the plasmoid instability is also a likely trigger for the onset of `fast' Hall/kinetic scale reconnection \cite{Huang2013,Daughton2009,Sheperd2010} (once the current sheet thickness drops below the ion {inertial length}) and so provides a bridge between macro-scales and micro-scales in large scale events. Stages (ii) and (iii) make the mechanism highly attractive for explaining the sudden onset of fast reconnection seen in solar flares or tokamaks.

The vast majority of previous work on the tearing instability has focused on the 2D problem. However, a few 3D studies have also been undertaken using simplified initial conditions of reduced dimensionality (using for example a Harris sheet equilibrium), usually including a strong guide field \cite[e.g.][]{Daughton2011,Baalrud2012}. An important consequence of introducing the third dimension, even in these simplified setups, is that when a guide field is present the islands formed in 2D become flux ropes -- loosely defined as regions of helical field -- with magnetic flux often wrapping multiple ropes \cite{lau1991,Daughton2011}. 

However, astrophysical magnetic fields such as those observed in the solar corona or planetary magnetospheres are typically highly complex in nature. In such complex magnetic fields current sheets may form preferentially at different topological features: 3D null points (isolated points at which the field strength is zero \cite{PriestTitov1996,LauFinn1990,Parnell1996}), separatrix surfaces, separator lines (field lines connecting two 3D nulls along the intersection of their separatrix surfaces \cite{PriestTitov1996,parnell2008}), and  Quasi-Separatrix Layers (locations at which the field line mapping has large but finite gradients \cite{PriestDemoulin1995,Aulanier2005}). Both the current sheets that form at these structures and the underlying magnetic fields are globally three-dimensional in nature. The question then arises: how does our understanding of tearing in 2D and ``guide field'' configurations translate to these more generalised geometries?

This question is particularly timely, as with the increase in computational resources current sheet tearing and fragmentation is now beginning to be observed in large scale numerical experiments; examples include experiments modeling Coronal Mass Ejections (CMEs) \cite{Karpen2012,Lynch2013}, coronal jets \cite{Moreno-Insertis2013} and flux emergence \cite[e.g.][]{MacTaggart2014}. Additionally, with the increase in spatial and temporal cadence of satellite observations, bright blobs of plasma thought to be the observational signatures of flux ropes/plasmoids are now regularly observed \cite[e.g.][]{Lin2009,Liu2010,Song2012}.

In this paper we take an important first step towards understanding 3D current sheet breakup by considering the fragmentation of current sheets formed at a 3D null point using some of the highest resolution simulations to date for such a dedicated set of experiments. In this investigation we focus on the general characteristics of the process -- how stable the layer is, what the overall dynamics are and how this affects the reconnection rate -- and plan to follow this with a second paper (hereafter referred to as Paper 2) giving a detailed account of how the magnetic topology evolves following the onset of tearing. As a contrast, we compare our results against an equivalent 2D setup initially containing a 2D null.

\begin{figure*}
\centering
\includegraphics[width=1.0\textwidth]{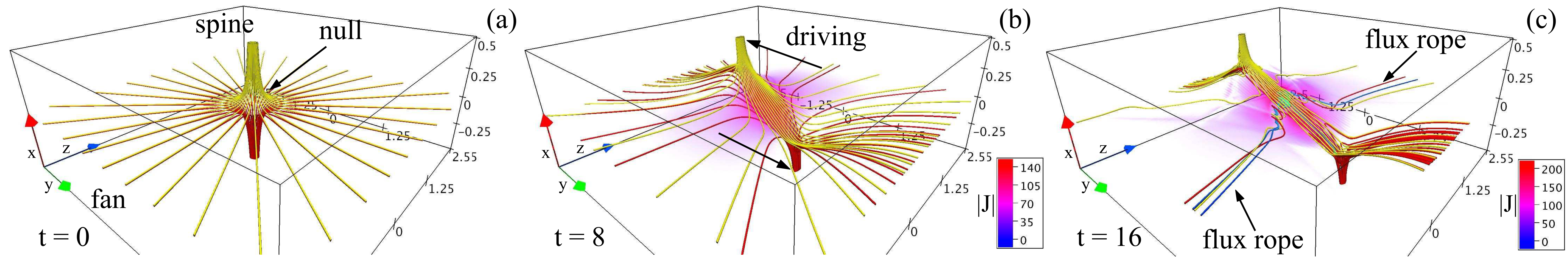}
\caption{(color online) 3D null simulation with $\eta=5\times 10^{-5}$ {showing the magnetic field at various times: (a) $t=0$, (b) $t=8$, (c) $t=16$}. Red/yellow: fieldlines traced {from rings of footpoints on the $x$-boundaries}. Blue: selected fieldlines within the flux ropes. Volume shading indicates current density.}
\label{fig:vap}
\end{figure*}

\section{Simulation Setup}
The simulation was carried out using the Copenhagen staggered mesh code \citep{Galsgaard1997}, solving the 3D MHD equations in the following non-dimensional form
\begin{eqnarray}
\frac{\partial \mathbf{B}}{\partial t} &=& \mathbf{\nabla}\times (\mathbf{v}\times\mathbf{B}) +\eta \mathbf{\nabla}^{2}\mathbf{B}\\
\frac{\partial(\rho \mathbf{v})}{\partial t} &=& -\mathbf{\nabla}\cdot(\rho \mathbf{v}\mathbf{v}) -\mathbf{\nabla}p  +\mathbf{j}\times\mathbf{B} \\
\frac{\partial \rho}{\partial t}&=& -\boldsymbol{\nabla} \cdot (\rho \mathbf{v}) \\
\frac{\partial e}{\partial t}&=& -\mathbf{\nabla}\cdot(e\mathbf{v})-p\mathbf{\nabla}\cdot\mathbf{v}+\eta j^2
\end{eqnarray}
where $\mathbf{J}=\mathbf{\nabla}\times \mathbf{B}$ is the electric current density, $\mathbf{v}$ plasma velocity, $\mathbf{B}$ magnetic field, $\rho$ density, $e$ thermal energy, $p=(\gamma-1)e$ gas pressure and $\eta$ the resistivity. The resistivity is set explicitly to a constant value throughout the volume, see Table \ref{table:runs} for values. Fourth-order hyper-viscosity terms (i.e. acting only upon hydro-dynamic quantities) are also included in the momentum and energy equations for numerical stability. Each simulation used a stretched grid, with points packed around the origin -- where the current sheet initially forms -- to properly resolve the current layer prior to tearing, discussed further below. Length and time units are non-dimensionalised such that one unit of time is the Alfv\'{e}n travel time across one unit of length in a uniform plasma and magnetic field with $\rho=1$ and $|{\bf B}|=1$. 
 
In the 3D simulations, an initial potential magnetic field containing a single radially-symmetric 3D null at the origin is formed by placing two magnetic point sources on the $x$-axis at $x=\pm 2.5$, outside the simulation volume of $[x,y,z]\in\pm[0.5,3.5,4]$, i.e.
\begin{equation}
{\bf B}(t=0)= \frac{B_0({\bf x}-{\bf x}_0)}{|{\bf x}-{\bf x}_0|^{3}}+\frac{B_0({\bf x}+{\bf x}_0)}{|{\bf x}+{\bf x}_0|^{3}},
\end{equation} 
where ${\bf x}=[x,y,z]^T$, ${\bf x}_0=[2.5,0,0]^T$. The strength of the sources is set to $B_0=2.5^{3}/2$, so that in the vicinity of the null the linearised field is given by $\mathbf{B} = [-2x,y,z]$. Similarly, in the 2D simulations the initial magnetic field was constructed using two line sources placed on the $x$-axis at $x=\pm 2.5$, outside a simulation volume of $[x,y]\in\pm[0.5,3.5]$, i.e.
\begin{equation}
{\bf B}(t=0)= \frac{B_0({\bf x}-{\bf x}_0)}{|{\bf x}-{\bf x}_0|^2}+\frac{B_0({\bf x}+{\bf x}_0)}{|{\bf x}+{\bf x}_0|^2},
\end{equation} 
where ${\bf x}=[x,y]^T$ and ${\bf x}_0=[2.5,0]^T$, with the strength of the sources set to $B_0=2.5^2/2$, giving a linearised field in the vicinity of the null of $\mathbf{B} = [-x,y]$.

The equilibrium is disturbed by two driving patches on the $x$-boundaries, centred on the $x$-axis and oppositely directed in $y$, {of the following form:
\begin{align}
\mathbf{v}(x= \pm 0.5) = \mp & \frac{A(t)}{4}\left[\tanh{\left(\frac{y-y_{0}}{l_{y}}\right)}-\tanh{\left(\frac{y+y_{0}}{l_{y}}\right)}\right] \nonumber\\
\times & \left[\tanh{\left(\frac{z-z_{0}}{l_{z}}\right)}-\tanh{\left(\frac{z+z_{0}}{l_{z}}\right)}\right] \hat{\mathbf{y}}
\end{align} 
with $y_{0}=2.1$, $z_{0}=0.5$, $l_{y}=0.3$ and $l_{z}=0.2$. In the 2D experiments, $z$ is set to zero in the above equation. $A(t)=0.1\tanh{(t)}$ ramps up the driving smoothly from zero to a constant speed, with an absolute value of $0.10$ in the centre of the patch} -- approximately $10\%$ of the local Alfv\'{e}n speed -- over a period of one time unit. The plasma in the volume is assumed to be an ideal gas ($\gamma=5/3$) and is initially at rest with $e=0.025$ and $\rho=1$. All boundaries are closed and line-tied (${\bf B}\cdot{\bf n}$ fixed, ${\bf v}={\bf 0}$ outside driving regions). Narrow damping layers are included on the boundaries to reduce the reflection of waves back into the volume.

\begin{table}[ht]
\centering 
\caption{Summary of simulations (${}^\dag$: signifies that the quantity has been multiplied by $10^{4}$).} 
\begin{tabular}{c c c c c c c} 
\hline\hline 
Case & $\eta^{\dag}$ & Resolution & $\Delta x^{\dag}_{min}$ & $\Delta y^{\dag}_{min}$ &  Unstable? \\ [0.5ex] 
\hline 
1 & $0.5$ & $[450,2000,200]$ & 7.4 & 28.7 & Yes\\ 
2 & $1$ & $[450,1000,200]$ & 7.4 & 57.3  & Yes\\
3 & $2$ & $[450,1000,200]$ & 7.4 & 57.3  & No\\
4 & $0.5$ & $[900,4000]$ & 3.7 & 14.3 & Yes\\
5 & $1$ & $[900,4000]$ & 3.7 & 14.3 & Yes\\ 
6 & $0.5$ & $[450,2000]$ & 7.4 & 28.7 & Yes\\
7 & $1$ & $[450,1000]$ & 7.4 & 57.3 & Yes\\ 
8 & $2$ & $[450,1000]$ & 7.4 & 57.3 & No\\
9 & $3$ & $[450,1000]$ & 7.4 & 57.3 & No\\
10 & $5$ & $[450,1000]$ & 7.4 & 57.3 & No\\
\hline 
\end{tabular}
\label{table:runs} 
\end{table}

\section{Threshold for Instability}

The field in the vicinity of the original null point is defined by spine and fan structures: the spine is defined by two field lines that asymptotically approach the null, and the fan is a continuous separatrix surface of field lines emanating outwards from the null which lies on the boundary of the two topological regions, Fig.~\ref{fig:vap}a. Once the driving begins the two spine footpoints are advected in opposite directions on the two driving boundaries. This disturbance in the field propagates into the volume, forming a current sheet localised to the weak field region around the null point as the spine and fan are brought close to one another by the action of the Lorentz force, Fig. \ref{fig:vap}b -- see also \cite{PontinBhatt2007,Wyper2012,mcLaughlin2008}. Flux then begins to reconnect across the spines and fan in a smooth quasi-steady manner \cite{Galsgaard2011}. The length (in $y$) and breadth (out of the driving plane: in $z$) of the current layer gradually increase due to a slight imbalance between the rate of reconnection in the sheet and the rate that flux is piled up at the edge of the layer by the driver. 

\subsection{Measured Quantities}
\begin{figure}
\centering
\includegraphics[width=0.47\textwidth]{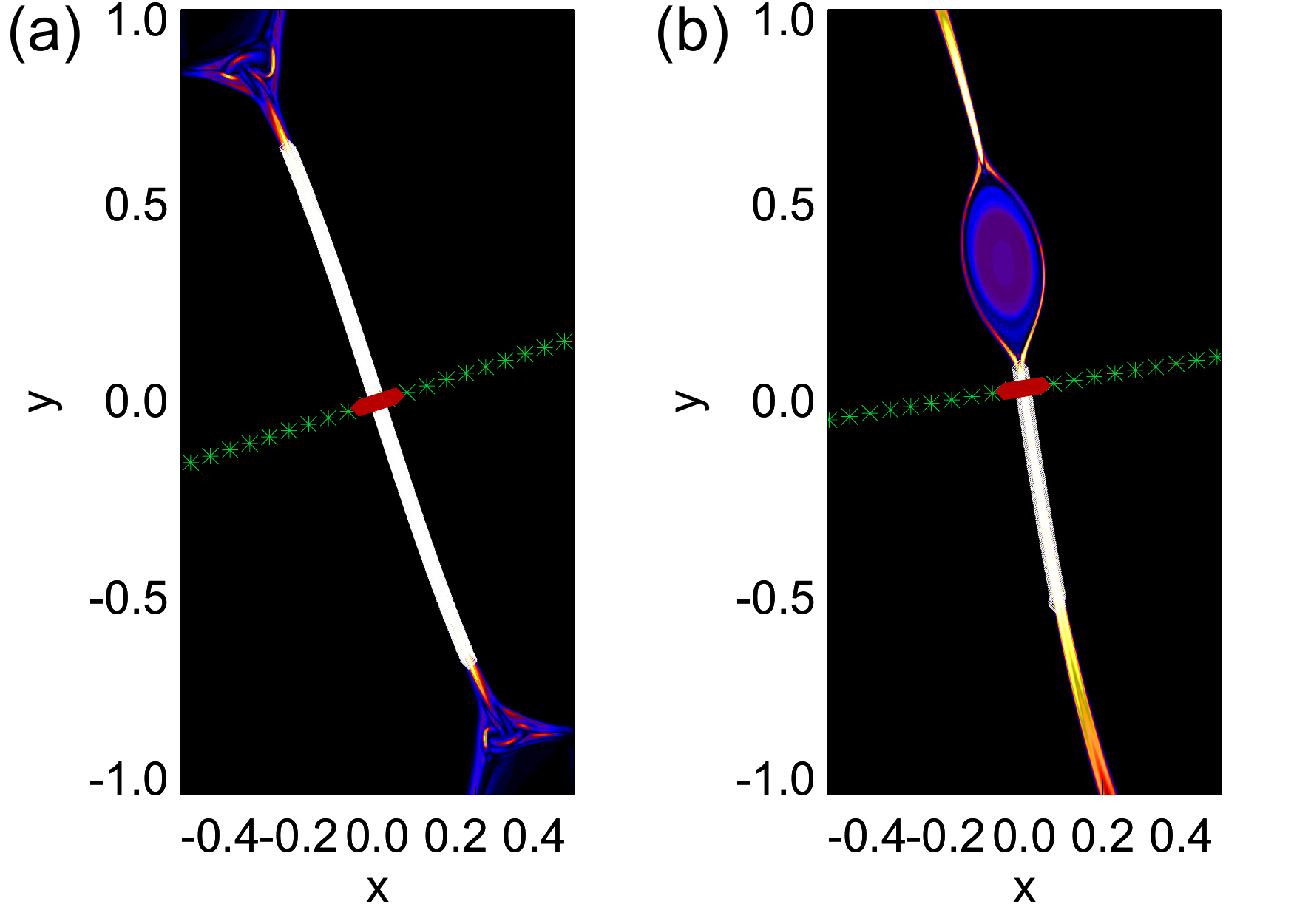}
\caption{(color online) Contour of $|\mathbf{J}|$ in a portion of the $xy$-plane overlayed with the positions of points used to determine  $S$ and $\mathcal{A}$: (a) 3D case 1 at $t=12$ -- prior to tearing. (b) 2D case 7 at $t=17$ -- after an island has formed, but before the layer becomes violently unstable. White points lie along the centre of the sheet where $|\mathbf{J}|\geq J_{max}/2$ (used to find $L^{*}$); green points indicate the direction of the locally perpendicular line passing through the site of $J_{max}$; red are the interpolated points used to find $\delta^{*}$ and $v_{a}$. The contours are scaled to half the maximum in each frame.}
\label{fig:res}
\end{figure}

Beyond a critical threshold the current sheet in some of the simulations then begins to fragment via the tearing instability, forming pairs of flux ropes -- helical regions of twisted field -- which snake across the current layer, Fig.~\ref{fig:vap}c. To determine the threshold of this tearing in the 3D experiments, $S=L^{*}v_{a}/\eta$ and $\mathcal{A}=L/\delta$ were found using the full width ($\delta^{*}=2\delta$) and length ($L^{*}=2L$) at half maximum of the sheet in the $xy$-plane, the plane of maximum spine-fan collapse. Due to the driving and the natural preference of the current layer to spread out across the fan separatrix surface, the current layer traces out a curved path in this plane \cite{PontinBhatt2007,Jorge2013}. To account for this, a method was developed to trace the relevant quantities along the current layer. First, the maximum value of $|\mathbf{J}|$ is identified within the layer, $J_{max}$. Starting at this location, a series of points following the curve of the current layer were found by stepping in both directions along the layer. This was continued until the values of $|{\bf J}|$ in each direction dropped below $J_{max}/2$. $L^{*}$ is then the distance along this curve between these two points. To obtain $\delta$ and $v_{a}$, $v_{a}$ and $|\mathbf{J}|$ are interpolated along another line of points, defined as the line which passes through the position of $J_{max}$ perpendicular to the current layer. $\delta$ is found from the distance between the half maximum points of $|\mathbf{J}|$ along this line, and $v_{a}$ from an average of the values at the edge of the current layer (where $|\mathbf{J}| \leq 0.01J_{max}$). Figure \ref{fig:res}a shows an example of the result of implementing this procedure. In some circumstances we also found it necessary to continue to measure these quantities once {an island/flux rope had formed}. In this case, the same procedure was applied to the current sheet containing the highest current in this plain -- see Fig.~\ref{fig:res}b.

We set two criteria for identifying the onset of the tearing instability in these simulations. The first is that due to the symmetry of the system, the first island/flux rope should form over the original, highly collapsed null point; therefore, for tearing to occur the null at the centre of the current layer must bifurcate. To check this we found the position of all nulls within the simulation volume using the trilinear method described in \citet{Haynes2007}. The second condition is that subsequent tearing should then occur following the formation/ejection of the first flux rope. If these two conditions are met for a given simulation then it is said to have reached the instability threshold around this time. The critical values of $S_{c}$ and $\mathcal{A}$ are then defined to be the values just prior to flux rope formation. As each is taken from a non-linearly varying experiment these values provide only an estimate for when the instability threshold is exceeded but do provide a basis for assessing the relative stabilities of the 3D and 2D setups.

\begin{SCfigure*}
\centering
\includegraphics[width=0.7\textwidth]{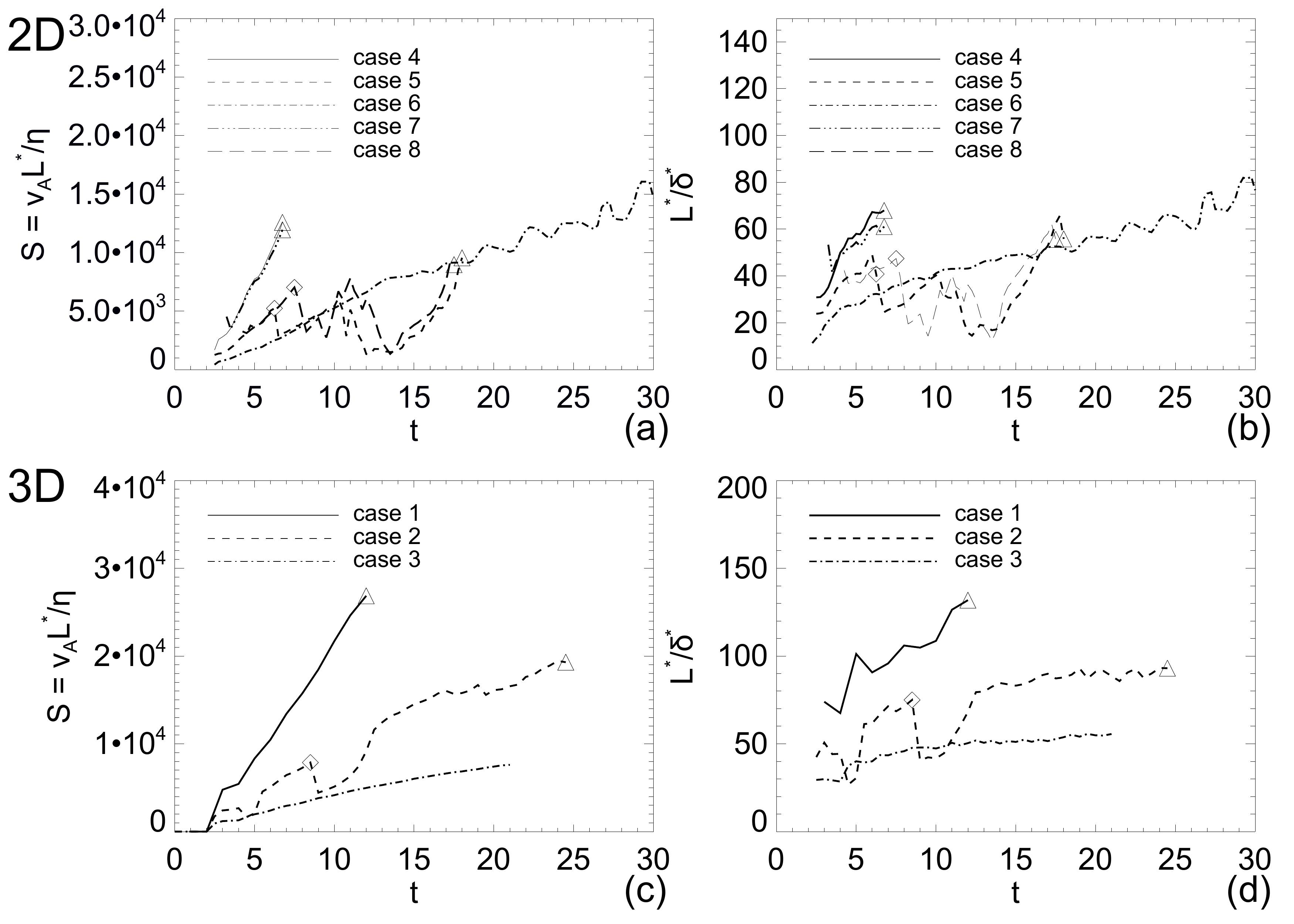}
\caption{Evolution of $\mathcal{A}$ and $S$ in the 2D (a,b) and 3D (c,d) simulations with different background $\eta$ values (see Table \ref{table:runs}). Triangles mark the null bifurcation that leads to the sheet becoming violently unstable. Diamonds denote the time a bifurcation occurs (producing an island/flux rope) but the flanking shortened current layers remain stable -- see text for details.}
\label{fig:arlund}
\end{SCfigure*}

\subsection{Results}
The speed and spatial profile of the driving flow were chosen to create a current layer which progressively lengthens, and to provide a quasi-steady inflow of flux towards the reconnection region. In theory, the current layer should lengthen until $S_{c}$ is reached (provided that the aspect ratio, $\mathcal{A}$ is high enough) enabling the value of $S_{c}$ to be obtained. However, unavoidable perturbations within the volume seem to trigger an early onset of tearing in some cases. Therefore, we measured $S$ and $\mathcal{A}$ beyond the initial island/flux rope formation in some of the experiments.

Figure \ref{fig:arlund} shows how $S$ and $\mathcal{A}$ change in time for cases 1-8. The measurement of the values is terminated when a simulation enters a highly fragmented, plasmoid/flux rope dominated phase of evolution. A current layer is deemed to have passed beyond the threshold for instability at this time. The values that $S$ and $\mathcal{A}$ take at this time are marked with triangles. Diamonds denote when a null bifurcation occurs (producing an island/flux rope which is subsequently ejected), but does not lead to further tearing in the flanking, shortened current sheet.

We describe first the results of the 2D simulations. For the runs with $\eta=5\times 10^{-5}$ and $1\times 10^{-4}$ (cases 4 and 5), the central $X$-point bifurcates and a magnetic island forms at $t\approx 8$. As mentioned, this is likely a result of a perturbation to the sheet caused by waves within the simulation volume (and also occurs in the higher resolution experiments -- cases 7 and 8). The ensuing evolution is slightly different in the two experiments. In case 4, the subsequent sheet thinning sets off further tearing in the flanking current layers before the island is ejected and the layer is deemed to have passed beyond the threshold for instability -- solid lines, Fig.~\ref{fig:arlund}(a,b). In case 5, the island grows large as it is slowly ejected, during which time the flanking current layers remain stable, and in fact take on a Petschek-like, opened out shape at this time -- shortening the dominant current sheet length ($L^{*}$). This is evident in the drop in both $S$ and $\mathcal{A}$ (as both are proportional to $L^{*}$), dashed lines, Fig.~\ref{fig:arlund}(a,b) -- $t\in [9,13]$. Beyond $t \approx 13$, the island has moved to near the exit of the main reconnection layer. The trailing current layer lengthens into a {Sweet-Parker-like} geometry once more (seen as an increase in $S$ and $\mathcal{A}$ -- $t\in [13,17]$, {see also Fig.~\ref{fig:res}(b)}) and evantually tears, forming several small islands and entering a plasmoid-dominated non-linear evolution at $t\approx 18$. The triple-dot dashed and long dash lines show the comparison with the respective high resolution experiments, which follow similar evolutions and shown an excellent agreement with the values of $S$ and $\mathcal{A}$ at which the layer becomes unstable. The run with $\eta=2\times 10^{-4}$ (case 8) remains stable throughout the experiment, but the current layer takes on a Petschek-like shape at later times $t \gtrsim 20$ which likely explains why the sheet remains stable despite attaining Lundquist numbers in excess of the other two runs. {Cases 9 and 10 also remain stable throughout their evolution, so have been omitted from Fig.~\ref{fig:arlund} in the interests of clarity}. From the values at which cases 4-5 and 7-8 became rapidly unstable, we conclude that for our 2D control setup the critical Lundquist number $S_{c} \approx 10^{4}$, occurring for aspect ratios above $\mathcal{A}_{min}\approx 50$. These values are broadly consistent with previous studies using less dynamically formed current sheets \cite[e.g.][]{Loureiro2005,Loureiro2007,Bhattacharjee2009}.

Turning now to the 3D simulations, we find that those with $\eta=5\times 10^{-5}$ and $1\times 10^{-4}$ (cases 1 and 2) form flux ropes (following a null bifurcation -- see below) and eventually descend into a highly fragmented, flux rope dominated state. In case 1 this occurs directly, with a flux rope pair forming at $t\approx 12$, leading to further rapid tearing and flux rope formation -- solid line, Fig.~\ref{fig:arlund}(c,d). In case 2, a null bifurcation occurs at $t\approx 8$ (diamonds, Fig.~\ref{fig:arlund}(c,d)) -- forming a flux rope pair. However, the main layer remains otherwise stable and recovers once the pair are ejected -- seen as a drop in $S$ and $\mathcal{A}$ between $t\in [8,12]$. The layer continues to lengthen until multiple flux ropes form at $t\approx 24$, where the layer is said to have passed the threshold for instability. Case 3 (with $\eta=2\times 10^{-4}$) remains stable throughout. The evolutions of cases 1 and 2 suggest that the threshold Lundquist number in the 3D experiments is around $S_{c}\approx 2\times 10^{4}$, occurring when the current layers have an aspect ratio of at least $\mathcal{A}_{min} \approx 100$. This suggests that 3D null current sheets are marginally more stable to tearing than 2D SP layers.

\subsection{Discussion Of Thresholds}
\begin{figure}
\centering
\includegraphics[width=0.48\textwidth]{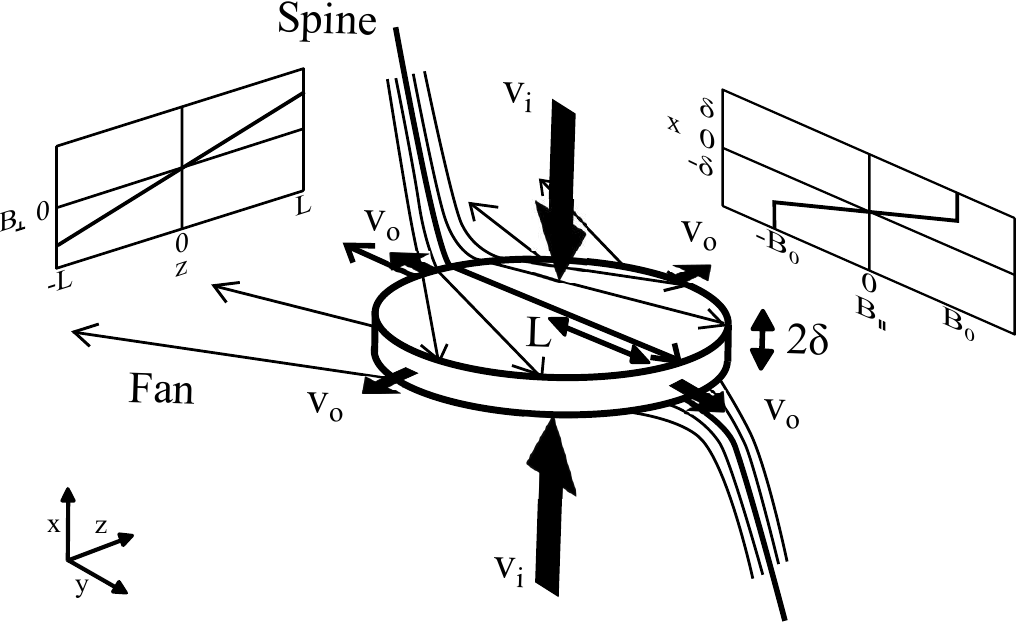}
\caption{Simplified model of the pre-tearing current sheet. $B_{\parallel}$ is the anti-parallel component of the field across the sheet at $B_\perp$ is the component perpendicular to the plane of null collapse. $v_i$ is the inflow velocity and $v_o$ the outflow velocity, both assumed uniform.}
%
\label{fig:cartoon}
\end{figure}

One reason why the instability threshold is marginally higher for the 3D null configuration may be that the plasma in the current layer is able to escape through the sides of the sheet. To demonstrate this, consider the diagram in Fig.~\ref{fig:cartoon}. This disk approximates the shape of the pre-tearing current layer in our simulations, Fig.~\ref{fig:vap}(b). Following a Sweet-Parker-type analysis, mass diffuses into the current sheet at a speed $v_{i}=\delta/\eta$, and mass continuity implies $2\pi L^2 v_{i} = 4\pi \delta L v_{o}$. Assumptions on the nature of the outflow then constrain the rate of flux transfer into the region. Choosing the simplest scenario of a radial outflow and assuming $B_{\perp}$ is passive within the layer, $v_{o}$ can be obtained (see \citet{Priest2002} for a similar example) by equating the inflow of free magnetic energy to the outflow kinetic energy: ${B_{0}}^2/2\mu = \rho {v_{o}}^2/2$. Combining these we find that:
\begin{equation}
v_{i}/v_{a} = \sqrt{2} S^{-1/2}, \quad \delta/L = S^{-1/2}/\sqrt{2} 
\label{eq:thresh1}
\end{equation}
where $S = L v_{a}/\eta$ and $v_{a}=B_{0}/\sqrt{\mu\rho}$ is the upstream Alfv\'{e}n speed based on $B_{\|}$. Eq.~\ref{eq:thresh1} shows that when plasma escapes radially through the sides of the sheet, the non-dimensional rate that flux is advected into the layer ($v_{i}/v_{a}$) is a factor of $\sqrt{2}$ faster, necessitating a thinner or longer sheet than in 2D. A radial outflow is a rather extreme assumption, given that the magnetic tension of newly reconnected fieldlines in the layer would be expected to launch plasma preferentially towards the ends of the sheet. Thus, the relationships above should be considered as upper bounds. It is interesting to note that \citet{Galsgaard2011} performed a series of 3D simulations similar to ours but restricted to the laminar stage of the evolution, i.e.~prior to any flux rope formation. They noted that in the quasi-steady reconnecting current sheet $Lv_i/\delta v_o\approx 1.5$ in a typical simulation, which lies between the 2D value of $1$ and the value of $2$ that would be obtained by combining the terms in Eq.~\ref{eq:thresh1}.

Returning to the discussion of thresholds, if we then further assume that the width of the 3D and 2D layers are comparable at the point of tearing ($\delta \approx \delta_{2D}$), then $L \approx \sqrt{2}L_{2D}$. Writing the 3D aspect ratio and critical Lundquist number in terms of its 2D counterpart then gives that:
\begin{equation}
\mathcal{A}_{3D} = \sqrt{2} \mathcal{A}_{2D}, \quad S_{c,3D} \approx \sqrt{2} S_{c, 2D}
\label{eq:scal}
\end{equation}
Given the number of assumptions made above, the relationships in Eq.~\ref{eq:scal} agree reasonably well with the simulation results. Line-tying on the $z$-boundaries (of the `out-of-plane component') may also act to inhibit the growth of the instability, although we do not believe that this will have a significant effect given that the tearing instability occurs initially in the symmetry plane, near which $B_z$ is weak.

\subsection{Resolution Of The Current Layer}
In these simulations, the current sheet forms at a time-varying angle relative the background grid (see also \cite{Galsgaard2011,PontinBhatt2007}), with the angle between the sheet and the $y$-axis reducing as the simulation progresses. To aid in fully resolving the layer, each simulation used a stretched grid with the majority of points packed around the $y$-axis, so that as each simulation progressed (and the current sheet aligned to the $y$-axis), the sheet became better resolved. The threshold of the plasmoid instability can be highly dependent upon the degree of numerical noise in a given numerical simulation \cite[e.g.][]{Huang2013}. To check the robustness of our results in the 2D setup, the runs which became tearing unstable (cases 6 and 7) were repeated with at least double the resolution (cases 4 and 5). The quasi-steady Sweet-Parker stage and the linear phase of the tearing instability agree closely, with small differences only arising in the non-linear tearing phase where the inter-island current layers can thin further at the higher resolutions. Given the large grid sizes, it was not practical to re-run the 3D experiments at higher resolutions, but each tearing unstable 3D experiment was checked to be sure that just prior to tearing there was a similar number of grid points spanning the current layer as in its 2D counterpart. As the sheet forms at an angle to the grid, this was found using $n_{equiv.}=n_{x}n_{y}/\left(n_{x}^2+n_{y}^2\right)^{1/2}$, where $n_{x}$ and $n_{y}$ are the number of points spanning the sheet in the $x$ and $y$ directions respectively.  A good resolution of the current sheet provides confidence that the explicit spatially-uniform resistivity used in all simulations is significantly larger than any numerical dissipation. Generally, $n_{equiv.}\geq 18$ between the edges of the current layer (where $|\mathbf{J}| \geq 0.01J_{max}$) which, combined with the low numerical dissipation afforded from the sixth-order spatial derivatives employed by the code, leads us to be confident that the analysis of the relative stability of the 2D and 3D setups uses simulations which are properly resolved.

\begin{figure}
\centering
\includegraphics[width=0.35\textwidth]{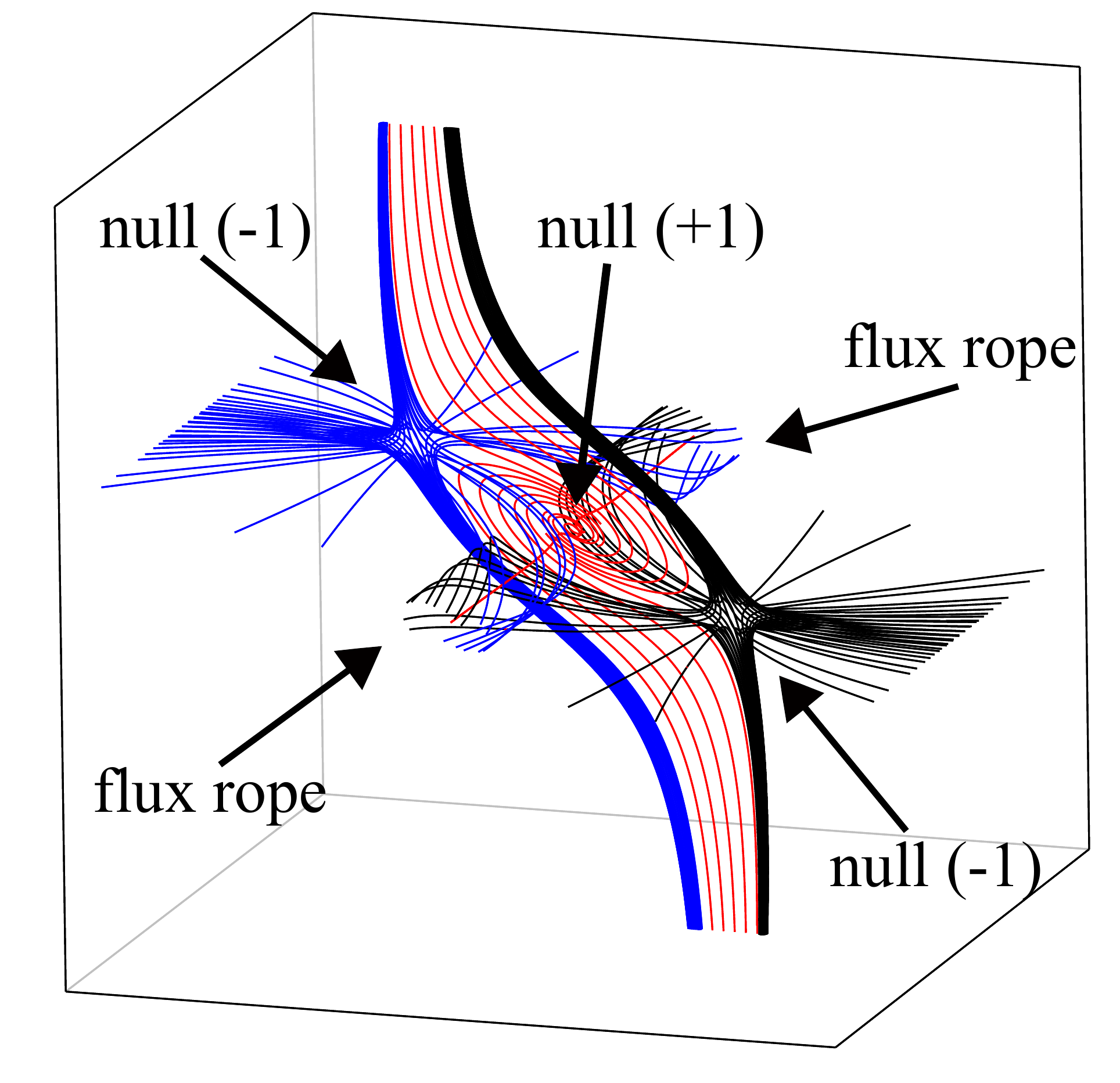}
\caption{A model magnetic field showing the magnetic topology following the bifurcation of the central 3D null within the current sheet. }
\label{fig:model}
\end{figure}

\begin{SCfigure*}
\centering
\includegraphics[width=0.7\textwidth]{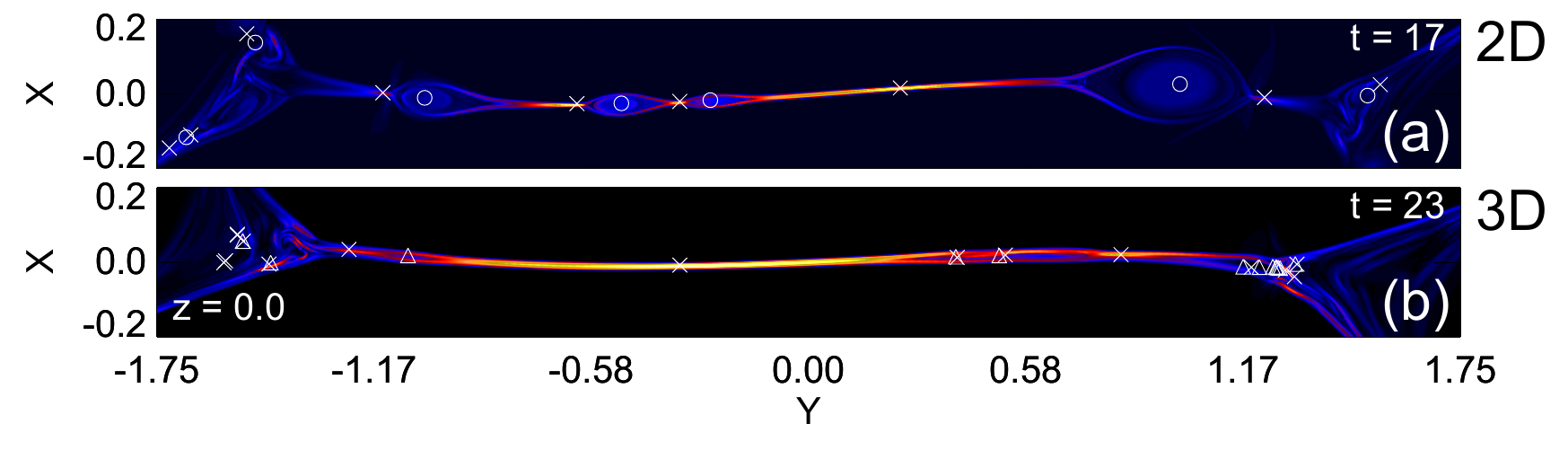}
\caption{(color online) Comparison of 2D islands with a slice through the 3D flux ropes. (a) $x$'s and $o$'s: X and O-points; (b) $x$'s and $\bigtriangleup$'s: 3D nulls within $z\pm 0.05$ with t.d. $-1$ and +1, respectively. Shading indicates $|\mathbf{J}|$, scaled to the maximum value. Each view has been rotated ($x,y \to X,Y$ by $12\,^{\circ}$ (a); $6\,^{\circ}$ (b)). Nulls were found using the trilinear method \cite{Haynes2007}.}
\label{fig:jpic}
\end{SCfigure*}

\begin{figure}
\centering
\includegraphics[width=0.48\textwidth]{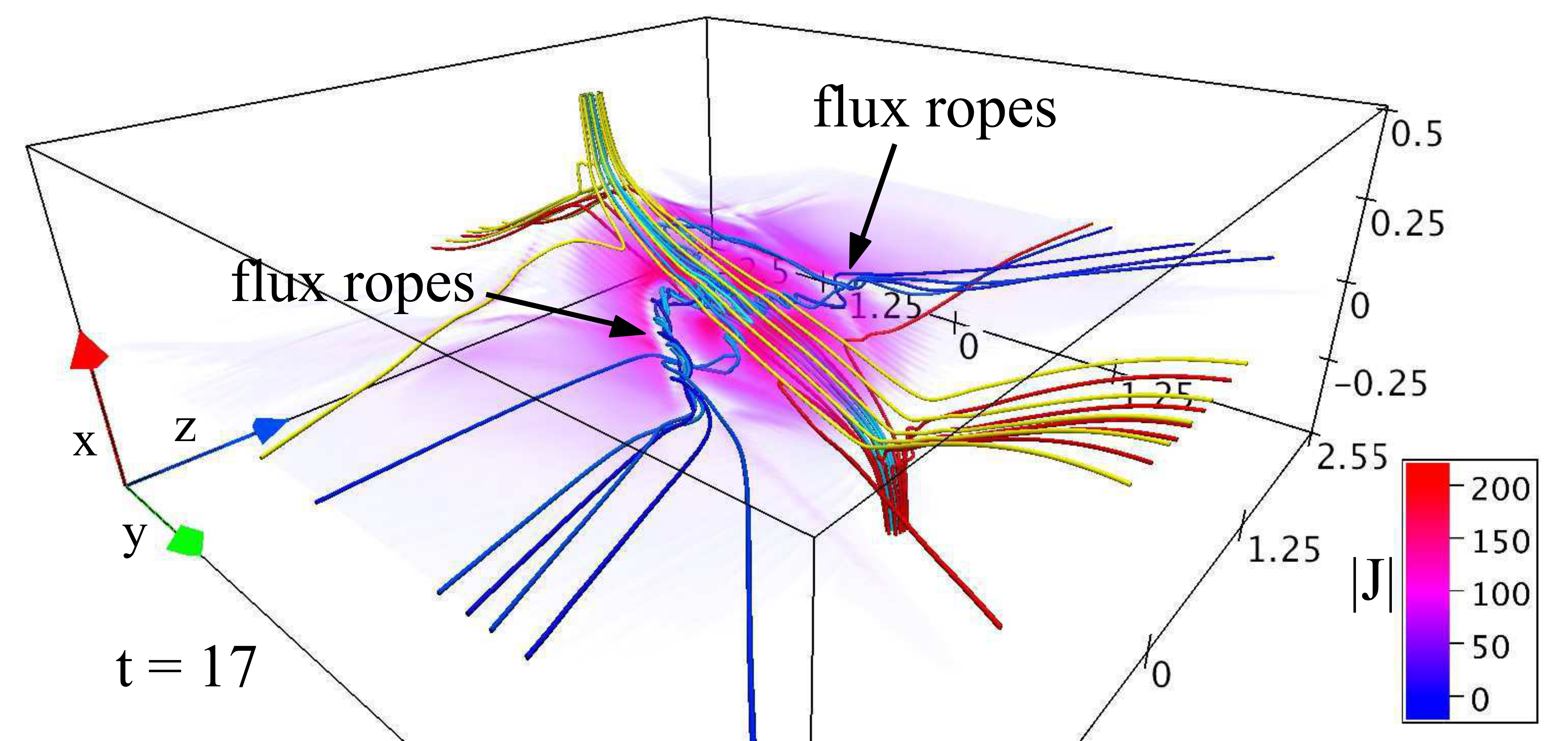}
\caption{(color online) Interaction of adjacent flux ropes following the ideal kinking instability. Blue fieldlines show four flux ropes intersecting in a turbulent-like region in the $xy$-plane. Yellow/red fieldlines show the global shape of the magnetic field.}
\label{fig:interaction}
\end{figure}

\section{Entering The Flux Rope Dominated Phase}
As in previous 3D studies of neutral sheets with guide fields, e.g. \citet{Daughton2011}, when the 3D null current sheet is unstable the non-linear evolution is dominated by interacting flux ropes. The first flux ropes form as a result of a bifurcation of the central 3D null point within the current layer. A detailed description of the evolving topology will be presented in Paper 2. Here we note that the magnetic configuration following the bifurcation of the original null point is as shown in Fig.~\ref{fig:model} (see also Fig.~\ref{fig:vap}c). The original central null, with a topological degree \citep{greene1993} (t.d.) of $-1$ undergoes a pitchfork bifurcation to produce a spiral null of t.d. $+1$ flanked by two nulls of t.d. $-1$. The newly formed spiral null sits at the intersection of two spiral field structures -- which we loosely designate as flux ropes. The crucial distinction that this 3D topology has from the closed islands formed by tearing in the 2D experiments is that the magnetic field within the flux ropes has an open structure due to the 3D nature of the field \cite{Schindler1988,lau1991}. As such, the tearing does not create distinct new (closed) topological regions; there remain throughout the evolution only two distinct flux domains.

The tearing which drives this bifurcation occurs over a finite patch of current sheet around the null. This launches torsional MHD waves along each respective rope, allowing the induced twist to propagate outwards. Additionally, plasma is permitted to flow outwards along each of the flux ropes. The associated mass and magnetic flux transport is likely the reason that the flux ropes in the 3D simulations have a much flatter cross section compared with the closed plasmoids observed in 2D, Fig.~\ref{fig:jpic}. Therefore, as a result of the 3D nature of the layer both the threshold for instability and the subsequent non-linear growth of the ropes differs from the 2D scenario.

Further differences between the 2D and 3D simulations arise as multiple ropes begin to form and evolve. Newly formed, highly twisted ropes appear to be unstable to an additional ideal instability \cite{Dahlburg2001,Dahlburg2002} which kinks them so that adjacent ropes interact, Fig.~\ref{fig:interaction} -- blue fieldlines. A stronger guide field is known to stabilise against this kinking in twisted flux tubes \cite{Dahlburg2003} but without a guide field component the instability results in a descent into turbulence \cite{Dahlburg2001}. We observe that the weak magnetic field in the center of the current layer is most susceptible to this kinking, leading to regions which exhibit a turbulent-like behaviour. Flanking these regions (away from the mid-plane, $z=0$) where the $B_z$ (`guide') field is stronger more coherent kinking flux ropes exist. The evolution of the field in these different regions is consistent with the idea that what we are observing is ideal kinking of the flux rope structures. Note that we have taken care to refer to the complex regions which form at the center of our layer as exhibiting a ``turbulent-like'' evolution as in our 3D simulations there is not sufficient resolution within each region to develop any kind of inertial range over which an energy cascade could occur. These regions clearly cannot with any confidence be deemed fully turbulent, but with greater resolution genuinely turbulent regions may form.

\begin{figure}
\centering
\includegraphics[width=0.5\textwidth]{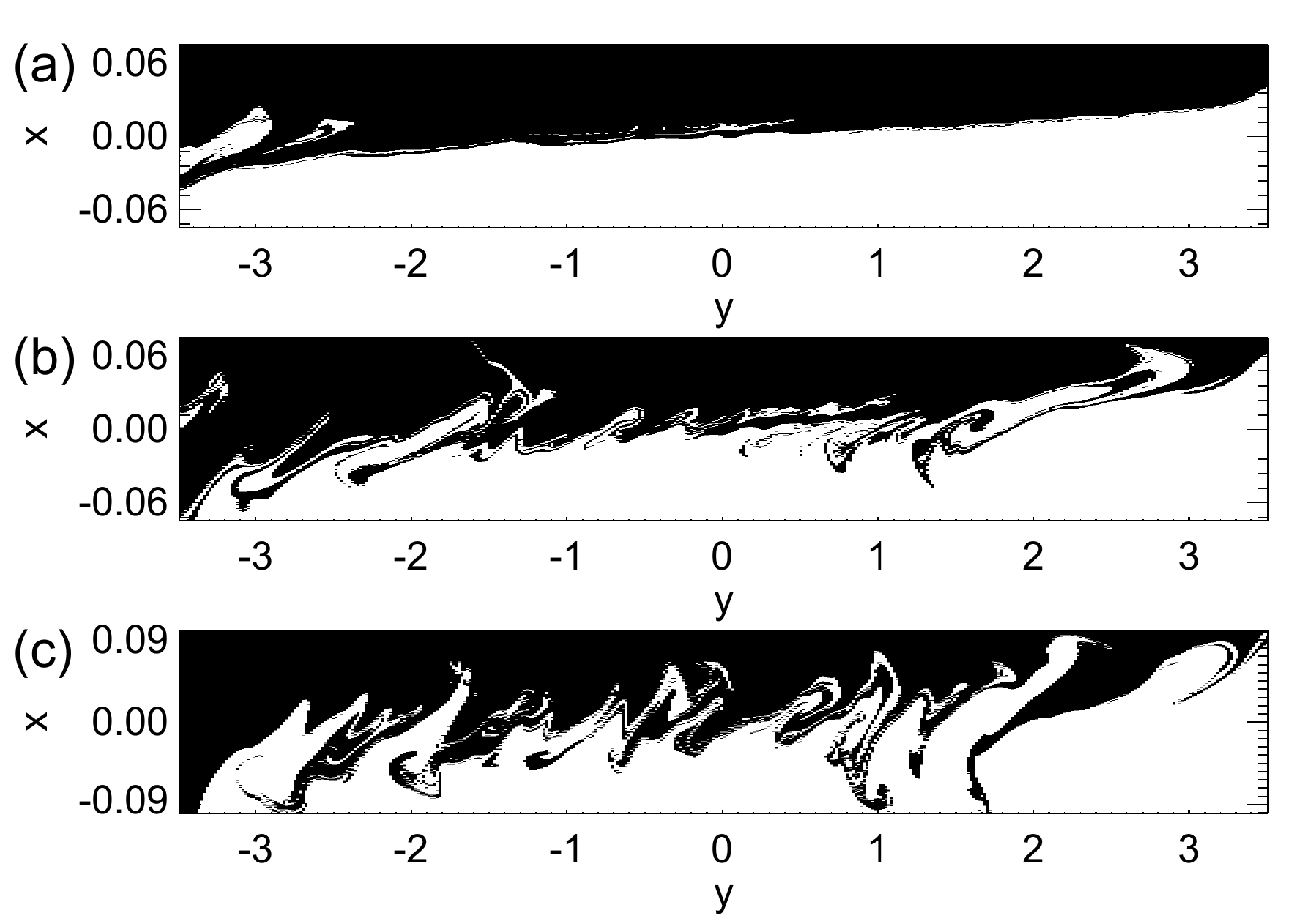}
\caption{Connectivity maps generated on the $z=-4$ boundary at (a) $t=17$, (b) $21.5$, and (c) $25$. Field lines from white points connect to the $x=-0.5$ boundary, those from black points connect to $x=+0.5$. (a) and (b) were produced using $80,000$ fieldlines, whereas (c) which is computed over a wider area used $160,000$.} 
\label{fig:maps}
\end{figure}

\section{Flux Mixing Across The Separatrix}
The original 3D null point field in our simulations partitions two regions of topologically distinct flux -- one where fieldlines have footpoints on the top ($x=0.5$) boundary, and the other with footpoints on the bottom ($x=-0.5$) boundary, Fig.~\ref{fig:vap}(a) -- red/yellow fieldlines. The separatrix surface intersects the side ($y$ and $z$) boundaries along a continuous line, {coincident with $x=0$ at $t=0$}. Once the driving begins and the current layer forms, flux is smoothly reconnected across the separatrix surface. This changes the identity of the separatrix footpoints on the side boundaries so that as the simulation progresses the curve along which the separatrix intersects the side boundaries becomes distorted. Due to the direction we have driven the spines, the separatrix moves upwards on the positive $y$ boundary ($y=3.5$), and downwards on the negative one ($y=-3.5$) -- see red/yellow fieldline evolution,  Fig.~\ref{fig:vap}(b,c). When the sheet fragments, on top of this general trend one would expect that the highly dynamic evolution of the field in the vicinity of the separatrix surface would lead to additional flux transport between these two topological domains.

In order to better understand how magnetic flux is mixed between the two domains, we produced a series of connectivity maps. These are formed by defining a grid of field line footpoints on each side boundary, tracing fieldlines from each point, and coloring each point according to whether the associated field line lies in the top or bottom domain. Field lines from white points connect to the $x=-0.5$ boundary, those from black points connect to $x=+0.5$. Figure \ref{fig:maps} shows connectivity maps at various times in case 1 taken from a side boundary ($z=-4$). The envelope within which flux is efficiently mixed presents as a thin boundary layer filled with extended spirals in the fieldline mapping. These spirals correspond to patches within the volume where the flux ropes twist up the separatrix surface and show that the magnetic field within these patches falls into distinct layers connected to the top and bottom boundaries. They also become progressively more complex as the simulation progresses -- indicative of the increased mixing of flux within the volume. Studying the evolution of these maps shows that these spirals form by wrapping up the fields of the two domains and subsequently relax again through an unwinding of the two layers. 

As is noticable from the red/yellow fieldlines in Fig.~\ref{fig:interaction}, the flux which threads into this thin boundary layer has a globally hyperbolic shape and connects with the top and bottom boundary within two small patches. The size of these patches is determined by an envelope of flux which just touches the edge of the boundary later -- within which all flux which threads the boundary layer at this time is contained. This envelope forms a hyperbolic structure of finite extent that mimics a spine-fan topology. 

Lastly, given the finite resolution of the maps, it is not clear whether the separatrix surface remains a smooth continuous surface or whether small, distinct flux domains form in this thin boundary layer. Recent work on the boundary between globally open and closed solar magnetic fields has shown that open field regions can bud-off and appear to be disconnected, but are in fact linked by a vanishingly thin line of flux, see \citet{Antiochos2007,Titov2011}. However, these investigations were in the context of genuinely global open and closed flux regions, so may not be directly applicable. What is clear is that if new topological domains are formed, they are constrained to exist within the thin boundary layer.

\begin{SCfigure*}
\centering
\includegraphics[width=0.7\textwidth]{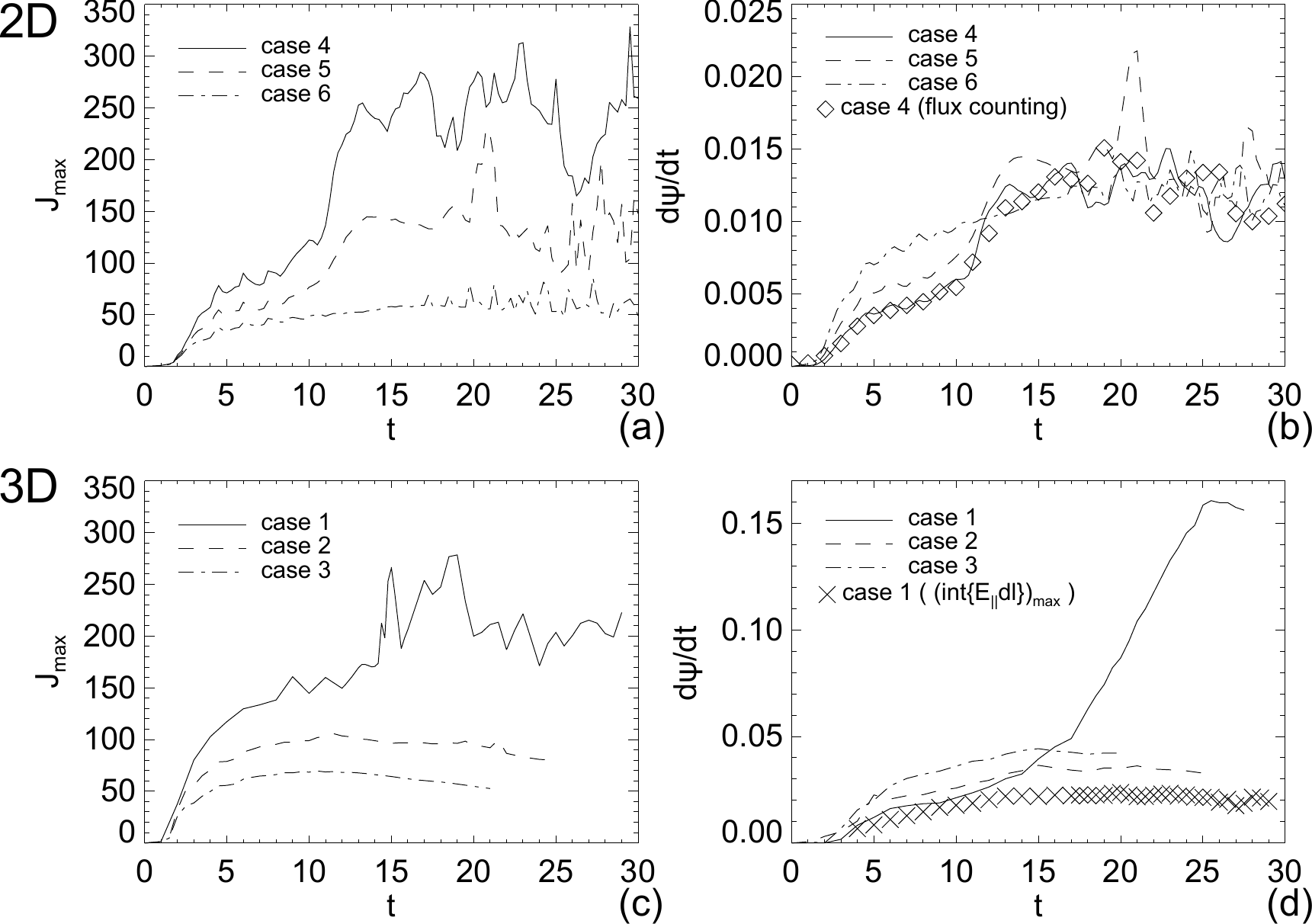}
\caption{(a), (c): maximum current in the volume. (b) $\eta J_{max} \approx$ reconnection rate in 2D, diamonds show a comparison with the flux counting method. (d) 3D reconnection rate obtained by the flux counting method. (e) comparison of $\left(\int{E_{\|}dl}\right)_{max}$ with the gross rate of transfer across the separatrix computed from the connectivity maps.}
\label{fig:jmrr}
\end{SCfigure*}

\section{Reconnection Rate}
Despite its distorted nature and the complexity of the field nearby, outside of the boundary layer there continues to remain only two distinct topological regions. It is clear by the evolution of the spirals in the fieldline mapping that the dynamics of the flux ropes within the volume is influential in transporting flux back and forth between these regions. This is in contrast to the 2D experiments whereby the magnetic islands are not directly involved in flux transfer between the different topological regions. In the 2D scenario the reconnection rate is simply given by the electric field at the dominant X-point -- a single X-point which lies at the intersection of the four global separatrix lines -- given by $\approx \eta J_{max}$. However in 3D it is known that reconnection occurs continuously throughout the non-ideal region, which is a layer of complex structure comprising the flux ropes and the fragmented inter-rope current sheets. The question then arises: how do we measure the reconnection rate in these simulations?

In general, for an isolated 3D reconnection region in the absence of null points the rate of flux transfer is given by the maximum of $\int{E_{\|} dl}$ along all fieldlines threading the region \cite{Hesse1988}. \citet{Pontin2005} showed that for a non-ideal region defined by a smooth current layer containing a single null  $\left(\int{E_{\|} dl}\right)_{max}$ measures the rate of flux transfer across half of the separatrix surface. However, once the current layer fragments multiple reconnection sites clearly form within the volume. {Not all have a sufficiently ideal region surrounding them for them to be considered as isolated.} \citet{Wyper2013b} considered a similar scenario to this where a current layer exists at a single null but is highly distorted -- leading to non-isolated patches of intense current within a large scale current layer. $\left(\int{E_{\|} dl}\right)_{max}$ in that case was shown to be of limited use for quantifying the rate at which flux is transferred between the two topological regions. A method relying upon an accurate knowledge of the position of the separatrix surface was presented which accounted for the multiple reconnection sites.

Since our separatrix surface becomes highly distorted, and as mentioned in the previous section difficult to identify at later times in these simulations, we use a different method to quantify the rate that flux is transferred between the two domains. The connectivity maps were used to apply the method of flux counting by comparing the maps at successive times and summing the number of fieldlines to have changed connectivity, weighted by the normal component of the field and the associated boundary area element. The finite temporal and spatial resolution of the maps mean that this provides a conservative estimate of the total flux transfer between the two topological regions.

Figure  \ref{fig:jmrr}(b,d) compares the temporal variation in the reconnection rate between the 2D (cases 6-8) and 3D (cases 1-3) experiments. The 2D runs exhibit a sharp increase in reconnection rate following the onset of tearing -- solid and dashed lines. In both the runs which became tearing unstable the rate at which this plateaus at is approximately the same, in agreement with the established theory that in the bursty non-linear phase of the plasmoid instability the average rate of reconnection becomes approximately independent of Lundquist number \cite[e.g.][]{Huang2013}. This is also true of the more highly resolved 2D experiments (cases 4 and 5). Although two of the 3D experiments become tearing unstable, computational constraints prevented us from running more than one far beyond this into the flux rope dominated regime. Figure ~\ref{fig:jmrr}(d), dashed and dot-dashed lines show that the rate of reconnection in cases 2 and 3 remains relatively steady once the current layer has formed ($t \geq 8$) and prior to tearing in case 2. By contrast, the rate of reconnection in case 1, which becomes tearing unstable early in the experiment, exhibits a smooth and substantial increase (approximately five-fold) following the onset of the tearing instability -- Fig.~\ref{fig:jmrr}(d) -- solid line. That the growth in reconnection rate is less explosive than in 2D we attribute to the fact that the enhancement of current in the layer after tearing (see Fig.~\ref{fig:jmrr}(a)) occurs only in small patches (Fig.~\ref{fig:interaction}), and therefore leads to only a small enhancement in the integral of $E_\|$ along any given fieldline ($E_\|=\eta J_\|$) threading the layer. This is demonstrated by the crosses in Figure  \ref{fig:jmrr}(d), which show that $\left(\int{E_{\|} dl}\right)_{max}$ does not undergo any significant enhancement once the tearing instability is sets in.

This substantial increase in reconnection rate occurs for very different reasons than in the 2D experiments. Whereas reconnection is sped up in the 2D experiments by shortening and thinning the current sheet at the dominant X-point, reconnection is sped up in the 3D runs by the introduction of many additional sites of flux transfer across the separatrix surface as a result of the fragmentation of the current layer. It is therefore unclear whether the scaling of the averaged rate of flux transfer in the later non-linear regime will follow that of the 2D scenario and become near independent of Lundquist number. Indeed, studies of current sheet fragmentation in magnetic braiding experiments \cite{Pontin2011b} and at separators \cite{parnell2008} hint that a decrease in $\eta$ leads to an increase in total amount of flux reconnected during a given event, facilitated by increased fragmentation and recursive reconnection. The scaling of the reconnection rate and cumulative reconnected flux with $\eta$ will be important quantities to explore in the future.

\section{Discussion}
In this study we have investigated how the tearing/plasmoid instability is triggered, and subsequently evolves, when both the underlying magnetic field and the current sheet are intrinsically three-dimensional. This was motivated by the fact that both observations and large scale simulations often contain complex three-dimensional fields within which current sheets form and fragment via a tearing-like process. We focused our attention upon how current sheets fragment when formed around a 3D null point, as 3D null reconnection is thought to be central to many astrophysical phenomena. By comparing with an equivalent 2D setup we showed that 3D null current sheets have similar stability properties to the 2D scenario (being marginally more stable), but that the subsequent dynamics exhibit a complex behaviour dominated by the formation, interaction and ejection of magnetic flux ropes. In particular, it was shown that an envelope with a global appearance of a spine-fan topology is created, within which flux between the two topological regions is efficiently mixed across the separatrix surface.

The findings of this work have implications on several areas of Heliophysics. In many applications, the fan separatrix surface of a pre-existing 3D null partitions regions of closed and globally open magnetic field. Within the context of the solar corona this occurs when a parasitic polarity emerges within a coronal hole \cite{Pariat2010,Titov2011,Moreno-Insertis2013}. Composition studies \cite[e.g.][]{Reames2013,Zurbuchen2007} have suggested that acceleration of both impulsive solar energetic particles and the slow solar wind may involve reconnection between open and closed magnetic flux -- often referred to as ``interchange reconnection'' \cite[e.g.][]{Edmondson2009}. We have shown that at Lundquist numbers typical of the solar corona ($S \approx 10^{14} \gg S_{c}$) the tearing of the current layer and the formation of flux ropes straddling the separatrix surface would lead to multiple sites on such a separatrix dome across which flux can be reconnected. This happens recursively between the open and closed fields and occurs within the mixed flux envelope with the global appearence of a spine-fan topology. As such our results could help to shed light on the origins of the slow solar wind.

We have also shown that following the onset of tearing new 3D magnetic null points appear along with the flux ropes, and that regions of turbulent-like field evolution occur as the flux ropes writhe and interact. Both turbulence and 3D nulls can be excellent particle accelerators {\cite[e.g.][and references therein]{Aschwanden2002,Stanier2012}}, and envisaging once more that our scenario is being played out at a coronal null point, the particle acceleration associated with these evolving structures may explain the anisotropic flare kernels observed at the separatrix footpoints of certain solar flares \cite[e.g.][]{Fletcher2001}. The finite width of the mixed flux envelope also provides a natural explanation for the finite width of SEP (Solar Energetic Particle) beams emitted in impulsive SEP events when the null configuration is such that one spine connects to open fields \cite[e.g.][]{Masson2012}.

\section{Conclusions}
This work was concerned with understanding the tearing instability in the context of 3D null point current sheets. Our main findings are that: 
\begin{enumerate}[(i)]
\item Current sheets that form about 3D null points are susceptible to an instability analogous to the plasmoid instability, but are marginally more stable than equivalent 2D neutral sheets. 
\item After the current layer tears a thin boundary layer is formed around the separatrix surface, within which flux from both topological domains is efficiently mixed. The flux threading this layer forms an envelope with a hyperbolic structure that mimics a spine-fan topology.
\item The mixing within this envelope leads to a substantial increase in the rate of reconnection between the two regions. 
\item The 3D evolution following tearing is dominated by interacting flux rope structures within the boundary layer. These interactions appear to be driven primarily by an ideal 3D instability which causes them to kink.
\item The flux ropes tend to have a much flatter aspect ratio in cross-section than the islands in an equivalent 2D simulation, since the tearing occurs in localised patches.
\end{enumerate}

{Looking ahead to future work, flux ropes and null points are fundamental elements of evolving magnetic fields at all scales throughout the heliosphere. In an effort to better understand how null points, flux ropes and reconnection are coupled in complex magnetic fields, we will follow this work with a second paper (Paper 2) giving a detailed description of the topology change and dynamics of the evolving post-tearing boundary layer straddling the separatrix surface. }

More generally, the initial magnetic field  {used} in our 3D simulations contained a radially symmetric 3D null. However, 3D nulls found in magnetic field extrapolations typically lack such radial symmetry \cite[e.g.][]{Masson2009}. The degree of null asymmetry has been shown to affect both how current sheets form at nulls and the subsequent reconnection process {\cite[e.g.][]{Galsgaard2011b,Cassak2007}}. Future work should be done to address how the degree of 3D null asymmetry affects the threshold for tearing to occur as well as the later flux rope dominated dynamics of the tearing mode and flux rope evolution. Additionally, work should be undertaken to address how the gross rate that flux is transferred across the separatrix surface scales with the diffusion parameters. An investigation of whether evolving patches of turbulence can be realised in 3D null current sheets would also be of great benefit, although this may require a different methodology than what is employed here.

Lastly, future work should also consider non-linear tearing during non-null reconnection in line-tied 3D magnetic fields, and the role that this plays in both the formation of the current layers and their subsequent dynamics. This is the subject of an ongoing investigation.

\begin{acknowledgements}
Financial support from the Leverhulme Trust and fruitful discussions with K. Galsgaard, G. Hornig and A. Haynes are gratefully acknowledged. DP also acknowledges financial support from the UK's STFC (grant number ST/K000993. Computations were carried out on the UKMHD consortium cluster funded by STFC and SRIF.
\end{acknowledgements}


%

\end{document}